\documentclass[12pt, letterpaper]{article}

\usepackage{amsmath}
\usepackage{amsfonts}
\usepackage{amssymb}
\usepackage{graphicx}
\usepackage{caption}
\usepackage{subcaption}
\usepackage{graphics,epsfig}
\usepackage{epsf}
\usepackage{tikz}
\usepackage[compat=1.1.0]{tikz-feynman}
\usepackage{epstopdf}
\usepackage{color}
\usepackage{braket}

\newcommand{\nc}{\newcommand}
\nc{\ba}{\begin{eqnarray}}
\nc{\ea}{\end{eqnarray}}
\newcommand\be{\begin{equation}}
\newcommand\ee{\end{equation}}

\def\bfk{{\bf k}}

\newcommand{\e}{{\bf{e}}}

\begin{document}
	
\vspace{0.5cm}
\begin{center}
		
\def\thefootnote{\fnsymbol{footnote}}
		
{\Large {\bf Schwinger Mechanism During Inflation}} 
\\[1cm]
		
{Soroush Shakeri $\footnote{s.shakeri@ipm.ir}$, 
Mohammad Ali Gorji$\footnote{gorji@ipm.ir}$, Hassan Firouzjahi$\footnote{firouz@ipm.ir }$,     
}
\\[0.5cm]

{\small \textit{School of Astronomy, 
Institute for Research in Fundamental Sciences (IPM) \\ 
P.~O.~Box 19395-5531, Tehran, Iran }} \\
\vspace{0.5cm}

\end{center}
	
\vspace{.8cm}
	
\begin{abstract}

We revisit the efficiency of Schwinger mechanism in creating  charged pairs during inflation. We consider a minimal setup of inflation in which the inflaton field is a complex scalar field charged under a $U(1)$ gauge field. There is a time dependent  conformal  coupling which pumps energy from the inflaton field to the gauge field to furnish a nearly constant background electric field energy density to drive the Schwinger mechanism. The coupling between the gauge field and the scalar field induces a time dependent effective mass for the inflaton field.  The requirement of a long period of slow-roll inflation  causes the  Schwinger mechanism to be highly  inefficient during inflation. The non-perturbative Schwinger mechanism can be relevant only towards the end of inflation and only on very small scales. This is in contrast to hypothetical models studied in literature in which the complex scalar field is a test field  and a constant electric field is imposed on the dS background by hand. We calculate the number of pairs of charged particles created perturbatively during inflation. We show that it is proportional to the amplitude of the quadrupolar statistical anisotropy and it is very small. Consequently, the back-reactions of created particles on magnetogenesis on large scales are negligible.

\end{abstract}
	
\vspace{0.5cm} \hrule
\def\thefootnote{\arabic{footnote}}
\setcounter{footnote}{0}
	
\newpage

\section{Introduction}

The time dependent nature of the background, whether originating from gravitation or electric fields, leads to  particle creation from vacuum \cite{2008LNP...738..193M}.  In cosmology, the presence of a time-dependent  background  in an expanding FLRW universe leads to the spontaneous creation of particles out of the vacuum \cite{1968PhRvL..21..562P},  while Hawking radiation is another example which is driven by the gravitational field in spacetimes with horizon such as  black holes \cite{Hawking:1974sw}. Moreover the existence of  a strong enough electric field can create pairs  of charged particles from vacuum by Schwinger mechanism \cite{1931ZPhy...69..742S}. 
Therefore, the investigation of  particle creation in the presence of both the gravitational and the electromagnetic (EM) fields is  important  for our
understanding of the evolution of the universe, from the early time up to the late-time stages \cite{2018PhRvD..98d5015F}.

Pair production phenomenon with  a background electric field in a de Sitter (dS) geometry  has been extensively discussed using  non-perturbative methods \cite{1994PhRvD..49.6327G,2014JHEP...10..166K} (see also \cite{Fischler:2014ama}), besides the  perturbative ones \cite{Cotaescu:2010bn}.  More specifically, the Schwinger effect in Minkowski spacetime is  a non-perturbative phenomenon which is exponentially suppressed below a critical electric field of about $1.3\times10^{18}\, \mathrm{V}/\mathrm{m} $, so the direct evidence of this process has not been observed yet.  This effect can be dynamically assisted by using  additional EM fields revealing  perturbative features of pair production \cite{PhysRevLett.103.170403}.  While  in  the flat spacetime  the perturbative computation of QED pair production has zero transition amplitude \cite{Weinberg:1995mt},  but the  perturbative pair production in dS in the presence of external EM fields is allowed.

The Schwinger effect in dS background has attracted much attentions in the past years.  In the case of 1+1-dimensional dS space,  Schwinger effect  revealed some quite different features thanks to the curvature of the spacetime \cite{1994PhRvD..49.6327G,Kim:2008xv}. It turns out that the current created by pair production increases as the electric field decreases which is different than our intuition of the  Schwinger effect in the flat space. The extension of the setup to the case of  3+1  dS background showed further unexpected results \cite{2014JHEP...10..166K,2016JCAP...07..010H,2018JCAP...10..023B}. It is shown that the Schwinger effect has the similar aspects even in D-dimensional dS spacetime \cite{PhysRevD.94.104011}.  However, recently it is claimed that  the unusual infrared (IR) behaviour (negative conductivity) of Schwinger current in dS background might be an artifact of regularization schemes \cite{2018JCAP...10..023B}. The created pairs during Schwinger process can back-react to the electromagnetic field in curved spacetime,  causing   some constraints on inflationary magnetogenesis models \cite{2014JHEP...10..166K, Sharma:2017eps, 2018PhRvD..98f3534S}. The created  pairs move along the electric field and produce current and then  change the conductivity of the ambient medium.  The Schwinger effect is also explored as a reheating mechanism in the context of a relaxion model for inflation \cite{Tangarife:2017rgl}.

Apart from the back-reaction  of Schwinger pairs to the background EM fields, the time dependent nature of the background electric field  in dS space has to be taken into account.  While the constancy of the energy density in a homogeneous field configuration  violates the second law of thermodynamics \cite{Giovannini:2018qbq},  most of the previous studies assume a constant electric field  to consider Schwinger  effect in dS  space. In order to develop a more realistic model one may consider Schwinger pair production during an inflationary era. In this framework, besides the quasi-dS geometry which is implied by inflation, we need an electric field which is present during inflationary period. In order to present such a setup we need to overcome several issues such as: i) The existence of background electric field and the associated induced charge current of Schwinger process manifestly break the de Sitter invariance of the background geometry, ii) Any vector field, like the electric field, is rapidly diluted because of the exponential expansion of the inflationary background. Providing appropriate setup in order to tackle these problems is the main goal of the present paper.
 
 The anisotropic inflation model provides a proper setup in which a persistent electric field can be  maintained during  inflation through a time-dependent gauge kinetic coupling, pumping energy continuously from the inflaton sector to the gauge field sector  \cite{2009PhRvL.102s1302W}.  As a result, one has an attractor solution  in which the energy density of the gauge field is a small and nearly constant fraction of the total energy density which can last for long enough period of inflation.

Recently several attempts were made in order to study Schwinger pair production by electric field coupled to inflaton and its back-reaction to the background geometry 
 \cite{2018PhRvD..98f3534S,2018JCAP...02..018G,2018PhRvD..98j3512K}.  This leads to some difficulties in solving the equations of motion of a charged scalar field  to find the mode functions and interpret them in terms of positive and negative frequency modes \cite{2018JCAP...02..018G}. Moreover, it is shown that the Schwinger effect during  inflation will cause an angular dependence on the primordial power spectrum and bispectrum \cite{2018arXiv181009815Z}.  In fact the charged particle production rate depends on the direction of these particle with respect to the background electric field and therefore leaves a unique angular dependence on the primordial spectra.  
 
 In all of these studies the complex scalar field,  which is responsible for the pair production,  is considered to be  a test field  during inflation while the dS spacetime is driven by the real inflaton field. In other words, the Schwinger mechanism of pair creation is decoupled from the inflationary sector. In addition, in most of these models, no mechanism is provided to generate the background constant electric field to drive the Schwinger mechanism.

 To address these shortcomings, in this work  we present a minimal setup  in which both the inflaton field and the charged scalar field are the same.  We  employ  the charged extension of the anisotropic inflation \cite{2009PhRvL.102s1302W} presented in  \cite{Emami:2010hy,Emami:2013bk} where the inflaton is charged under a $U(1)$ gauge field. At the background level, the model is physically the same as the anisotropic inflation. In this model  the inflaton field drives the quasi-de Sitter expansion through the slow-roll conditions while simultaneously pumping energy into the gauge field  sector to furnish a nearly constant background electric field.  At the level of perturbations, the pairs are naturally produced from the quantum fluctuations of the complex inflaton field.  We  show that the conventional nonperturbative Schwinger effect  is important only at the late stages of inflation.  Since in most of the inflationary era the electric field energy density is required to be small to allow slow-roll inflation,  pair production can only happen perturbatively. Note that the particle creation by a cosmological anisotropic Bianchi I universe in the presence of a constant electric field has been considered before in \cite{Villalba:1999aa}.

Since the quantum fluctuations of the inflaton field also generate the curvature perturbations \cite{Emami:2013bk}, the parameters space of the model are tightly restricted by the CMB observations \cite{2014JCAP...08..027C}.  Moreover, the anisotropic inflation model predicts  quadrupolar statistical anisotropies in the CMB angular power spectrum which are highly constrained by CMB observations \cite{Ade:2015lrj}.  Therefore some of the unusual aspects of the Schwinger mechanism in dS space which were obtained in previous studies (as summarized above) may be the results of a large and unrealistic  parameter space in which  no links between observations and theoretical results were made.

The rest of the paper is organized as follows. In Section \ref{model} we review the model of charged anisotropic inflation and explain how a nearly constant electric field in quasi-de Sitter background arises  in this model.  In Section \ref{pert} the perturbation analysis of the model are presented.  In Section \ref{pair} we compute the charged pair production rate while in Section \ref{power} the power spectra of the curvature and isocurvature modes are presented, followed by summaries and discussions in Section \ref{discussion}.


\section{The Model}\label{model}

In order to study the Schwinger process during inflation, we consider the model studied in \cite{Emami:2010hy}, containing a charged complex scalar field as the  inflaton  field and a $U(1)$ gauge field $A_{\mu}$,  
\begin{eqnarray}\label{action}
S=\int d^{4}x \sqrt{-g} \Big[
\frac{M_{P}^{2}}{2}R - \frac{1}{2}D_{\mu}\varphi \overline{D^{\mu}\varphi}
-\frac{1}{2}m^{2}\varphi \bar{\varphi}-\frac{1}{4} 
f^{2}(\varphi,{\bar \varphi}) F_{\mu \nu}F^{\mu \nu}\bigg] \,,
\end{eqnarray}
where the field strength tensor $F_{\mu\nu}=\partial_{\mu}A_{\nu}-\partial_{\nu}A_{\mu}$ and the covariant derivative is given by $D_{\mu}=\partial_{\mu}+i\mathbf{e}A_{\mu}$ in which ${\bf e}$ is 
the gauge coupling constant. This model is an extension of the anisotropic inflation model \cite{2009PhRvL.102s1302W} to the case of complex inflaton field where its perturbation analysis were  studied  in  \cite{Emami:2013bk, 2014JCAP...08..027C}. The isotropic version of the model is also 
recently proposed in Ref. \cite{2018arXiv181207464F} where the inflaton is charged under a triplet 
of $U(1)$ gauge fields. 

Here we briefly review the main results in this setup at the background level which are studied in details in \cite{Emami:2010hy,Emami:2013bk, 2014JCAP...08..027C}. 

In the model (\ref{action}) the gauge kinetic coupling $f$ breaks the conformal invariance such that the background gauge field survives the exponential expansion.  While this coupling  represents a non-renormalizable interaction, this  kind of term is usually  used for inflationary model buildings in  low energy expansions in the spirit of the effective field theory.  For the sake of simplicity, we
assume an axially symmetric structure in field space and therefore the coupling function becomes only a function of the amplitude of the complex field $\varphi \bar \varphi =|\varphi|^2$ so 
$f(\varphi,{\bar \varphi}) = f(|\varphi|)$. Taking this assumption into account, it is better to write the complex field in the polar coordinates as
\begin{eqnarray}\label{phi-polar}
\varphi = \rho \, e^{i\theta} \,,
\end{eqnarray}
where $\rho$ represents the amplitude while $\theta$ is the phase. Working with the above polar coordinates makes the calculations simpler. The coupling function then is only a function of the radial coordinates, $f(|\varphi|)=f(\rho)$.

Varying the action (\ref{action}) with respect to the gauge field $A_{\mu}$, we find the Maxwell equations
\begin{eqnarray}\label{Maxwell}
\nabla_{\mu} \big( \sqrt{-g}f^2F^{\mu\nu} \big) = \mathbf{e}J^{\nu} \,,
\end{eqnarray}
where we have defined the $4$-current as 
\begin{eqnarray}\label{J}
J^{\nu}\equiv \rho^2 \sqrt{-g}(\partial^{\nu}\theta+\mathbf{e}A^{\nu}) \,,
\end{eqnarray}
which  satisfies the continuity equation $\nabla_{\mu}J^{\mu}=0$. 

The existence of a background gauge field clearly breaks the isotropy and we have to consider a Bianchi spacetime for the background geometry. However, when studying the cosmological perturbations,  we can neglect the effects of the anisotropy in geometry as far as the size of anisotropy is sufficiently small. The statistical anisotropies in cosmological observables are predominantly induced from the matter (electric field) perturbations \cite{ Emami:2013bk, 2014JCAP...08..027C}.  Therefore, we consider the isotropic FLRW background geometry
\begin{eqnarray}\label{FRW-t}
ds^{2} = -dt^{2} +a(t)^{2} \delta_{ij}dx^{i}dx^{j} \,,
\end{eqnarray}
where $a(t)$ is the scale factor. For the gauge field, we take the following time-dependent configuration
\begin{eqnarray}\label{At}
A_{\mu} = ( 0, A(t), 0, 0) \,,
\end{eqnarray}

which preserves homogeneity while  breaking the isotropy.

Using Eqs.  (\ref{phi-polar}) and  (\ref{At}), the time component of the Maxwell equation (\ref{Maxwell}) implies ${\dot \theta}=0$ which shows that the phase $\theta$ does not play any role at the level of background. This is the advantage of working with the polar variables $\rho$ and $\theta$ rather than the original fields $\varphi$ and $\bar{\varphi}$.

The spatial components of the Maxwell equations (\ref{Maxwell}) yields 
\begin{eqnarray}\label{Max-i}
\partial_{t}(f^2 a \dot A ) = - \mathbf{e}^2 a \rho^2  A\,. 
\end{eqnarray}
In comparison with the standard anisotropic inflation \cite{2009PhRvL.102s1302W}, the right hand side of the Maxwell equation has the nonzero current coming from the induced mass term by the gauge field.  This  term also induces an effective mass $\mathbf{e}^2 A_\mu A^\mu$ for the inflaton field. However, in order not to destroy the slow-roll condition, i.e. the inflaton mass to be small compared to the Hubble expansion rate during inflation,   the induced mass term should be negligible during most of the period of inflation \cite{Emami:2013bk}. The background is therefore similar to the case of standard anisotropic inflation setup while interesting features of the models with the effects of $\mathbf{e}^2$ appear at the level of perturbations  \cite{2014JCAP...08..027C}. 

The $0-0$ component of the Einstein equations yields the Friedmann equation
\begin{eqnarray}\label{Friedmann}
3M_{P}^2H^2 = 
\frac{1}{2} \dot{\rho}^2 + \frac{1}{2}m^2\rho^2
+ \frac{1}{2}a^{-2}f^2\dot{A}^2 + \frac{1}{2} {\bf e}^2 a^{-2} \rho^2 A^2 \,,
\end{eqnarray}
where $H(t) = {\dot a}/a$ is the Hubble parameter and  a dot indicates the derivative with respect to the cosmic time. Combining the above result with the $i-i$ component of the Einstein equations, we find the time evolution of the  Hubble parameter as
\begin{eqnarray}\label{Raychuadhuri}
M_{P}^2\dot{H} = -\frac{1}{2} \dot{\rho}^2 - \frac{1}{3}a^{-2}f^2\dot{A}^2  \,.
\end{eqnarray}
Since the charged mass term $ \frac{1}{2} {\bf e}^2 a^{-2} \rho^2 A^2$ behaves as a potential term giving a mass to the inflaton field, it did not appear in the time evolution of the Hubble parameter.

The right hand side of (\ref{Friedmann}) is the total energy density where the first two terms are the standard energy density of the inflaton. The third term is the kinetic energy of the gauge field coming from the coupling between the inflaton and the Maxwell terms while the last term comes from the charged mass  term which, as we discussed above, is only important at the final stages of inflation. As soon as the induced mass term by the gauge field ($\mathbf{e}^2 A_\mu A^\mu \rho^2$)  dominates,  it terminates inflation quickly. 

Taking the variation with respect to $\varphi$, we obtain the Klein-Gordon equation as 
\begin{eqnarray}\label{KGE}
\ddot{\rho}+3H\dot{\rho}+m^2 \rho 
= a^{-2} f f_{,\rho}\dot{A}^2-\mathbf{e}^2a^{-2}\rho A^2 \,,
\end{eqnarray}
where $f_{,\rho} = \partial_{\rho}f$. The above equation shows the direct coupling between the inflaton and gauge field in this model. 

Our model describe two different phases: i) inflationary stage when the effects of charged mass term is negligible and the setup is very similar to the standard anisotropic inflation model containing a real inflaton field  \cite{2009PhRvL.102s1302W}.  ii) When the charged mass term dominates, for instance the last term  in Eq. (\ref{Friedmann}),  so it quickly terminates slow-roll inflation. During the first stage, which is the most of the period of inflation, we can neglect the charged mass term in all the background equations (\ref{Max-i}), (\ref{Friedmann}), and (\ref{KGE}).

Now, we should choose an explicit functional form for the coupling function $f(\rho)$. In the case of coupling of axion with gauge fields, the desired symmetry of the system under consideration
uniquely fixes the functional form so that the coupling function is linearly proportional to the
axion (see Refs. \cite{Peloso:2016gqs}). Here, however, we deal with a phenomenological 
inflationary scenario and, as shown in \cite{2009PhRvL.102s1302W}, with an appropriate form of the conformal coupling $f(\rho)$, the gauge field drags energy continuously from the inflaton sector which prevents the dilution of the vector field in the exponentially expanding universe. This is an attractor solution in which the electric field energy density reaches a small but a nearly constant fraction of the total energy density. Let us elaborate more on this effect. During the most period of inflation the effects of the induced mass term $\mathbf{e}^2 A_\mu A^\mu \rho^2$ is negligible so we can easily integrate the Maxwell equation (\ref{Max-i}) obtaining 
\begin{eqnarray}\label{Adot}
\dot{A} = \frac{q_{0}}{a f^2} \,,
\end{eqnarray}
where $q_0$ is an integration constant.  

Note that the functional form of the gauge coupling function  $f(\rho)$ determines the time dependence of the electric field in the model. The third term in the right hand side of Friedmann equation (\ref{Friedmann}) is the  energy density of the vector  field  which  in the case of $f=1$  decays like $a^{-4}$. In order to prevent this dilution, using Eq.  (\ref{Adot}),  it can be seen that if we choose $f\propto {a}^{-2}$, the energy density of the vector field remains constant. This result at background level determines  the functional form of $f(\rho)$ \cite{2009PhRvL.102s1302W}
 \begin{eqnarray}\label{f-phi}
 f(\rho) = \exp{ \Big( \frac{c \rho^{2}}{2M_{P}^{2}} \Big) } \,,
 \end{eqnarray}
where $c\ge 1$ is a parameter. With this form of $f(\rho)$, the system reaches the attractor regime in which the gauge field's energy density becomes a constant fraction of the total energy density \cite{2009PhRvL.102s1302W}. 
 
 Alternatively, the time-dependence of $f(\rho)$ can be written as 
 \begin{eqnarray}\label{f-c}
f = \Big( \frac{\tau}{\tau_e} \Big)^{2c} \,,
\end{eqnarray}
where  $\tau$ is the conformal time defined as $\tau = \int {dt/a(t)} $ and $\tau_e$ denotes the time of end of inflation. 

Substituting the above result, the energy density of the vector field turns out to be
\begin{eqnarray}
\rho_{E} \equiv \frac{1}{2} a^{-2} f^{2} \dot{A}^2 =
\frac{1}{2} q _{0}^{2} H^4 \tau_e^{4c} \tau^{4(1-c)} \,.
\end{eqnarray}
From the above relation we see that the energy density of the vector field is almost constant during the inflation. Now, it is easy to  interpret the integration constant
$q_0$. Demanding $\rho_{E}|_{\tau=\tau_e} = E_0^2/2$, we find 
\be\label{E0}
q_0 = \frac{E_0}{H^2\tau_e^2} \,,
\ee
where $E_0$ is the amplitude of the electric field at the end of inflation. Substituting Eqs. (\ref{f-c}) and (\ref{E0}) in Eq. (\ref{Adot}) and then integrating, we find that
\begin{eqnarray}\label{A-c}
A(\tau) = \frac{1}{-4c+1} \frac{E_{0}}{H^2\tau} \Big(\frac{\tau_e}{\tau}\Big)^{4c-2} \,.
\end{eqnarray}
During the final stages of inflation when $ \tau \sim \tau_e$   we find $A(\tau) \approx -\frac{E_0}{H^2\tau}$ which is the same as the ansatz supposed in  \cite{2014JHEP...10..166K}. However, note the important difference that during most of period of inflation $A(\tau)$ scales like $A(\tau) \sim \tau^{-3}$ which is quite different from the ansatz employed in  \cite{2014JHEP...10..166K} and in other works dealing with setup similar to \cite{2014JHEP...10..166K}. This is the key difference which significantly reduces the efficiency of the Schwinger mechanism during inflation. 

During the attractor phase,  the contribution of the gauge field to the total energy density of the model is given by the ratio \cite{2009PhRvL.102s1302W}
\begin{eqnarray}\label{R}
R\equiv\frac{\rho_{E}}{\rho_{\phi}} = \frac{E_{0}^2f^{-2}a^{-4}}{m^2\rho^2+\dot\rho^2} 
= \frac{c-1}{2c}\epsilon=\frac{I}{2}\epsilon \,,
\end{eqnarray}
where  we have introduced the anisotropy parameter $I\equiv(c-1)/c$ and $\epsilon$ is the slow-roll parameter which is given by
\begin{eqnarray}
\epsilon = -\frac{\dot H}{H^2} = \frac{1}{2c} \Big(\frac{V_{,\phi}}{V}\Big)^2 \,.
\end{eqnarray}

It is worth mentioning that there are no limitations on the values of $c$ and $I$ at the background level except that  $c\geqslant1$. However, at the perturbation level, in order to satisfy the observational constraints  on CMB anisotropies \cite{Ade:2015lrj} one requires that  $I \lesssim 10^{-7}$ \cite{Emami:2013bk,Kanno:2010ab, Abolhasani:2013zya}.

\begin{figure}
	\caption{The phase space plots of ($\rho,\dot \rho$)}
	\begin{subfigure}[b]{0.495\textwidth}
		\includegraphics[width=\textwidth]{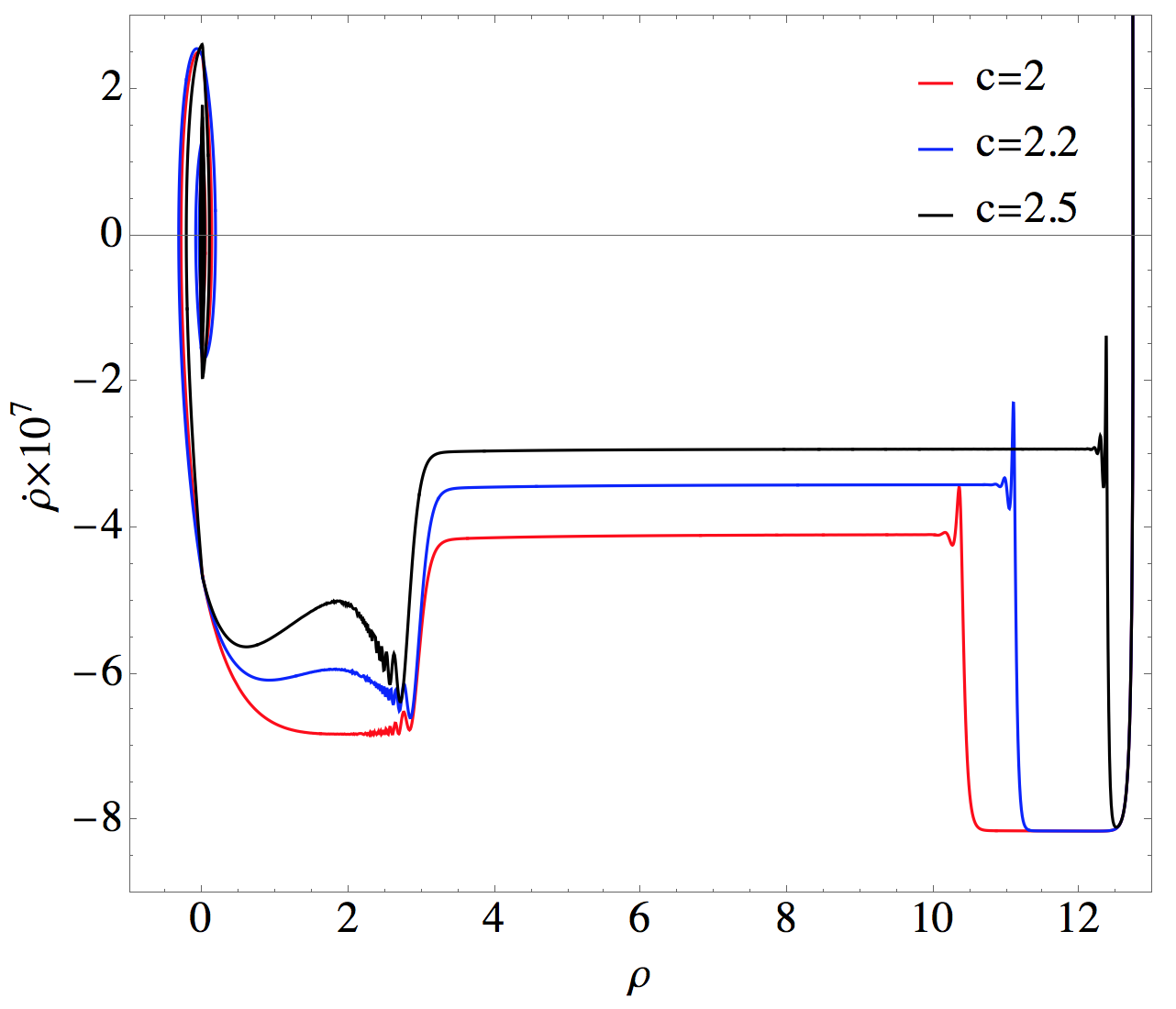}
		\caption{The phase space plot  with $\mathbf{e}=0.01$ for different values of 
		$c=2,2.2$ and $2.5$ from top to bottom, respectively. We have also set
		$m=10^{-6}M_{P}, \rho(0)=12M_{P}$ and $\dot \rho(0)=0$.}
		\label{cf1}
	\end{subfigure}
	\hfill
	\begin{subfigure}[b]{0.49\textwidth}
		\includegraphics[width=\textwidth]{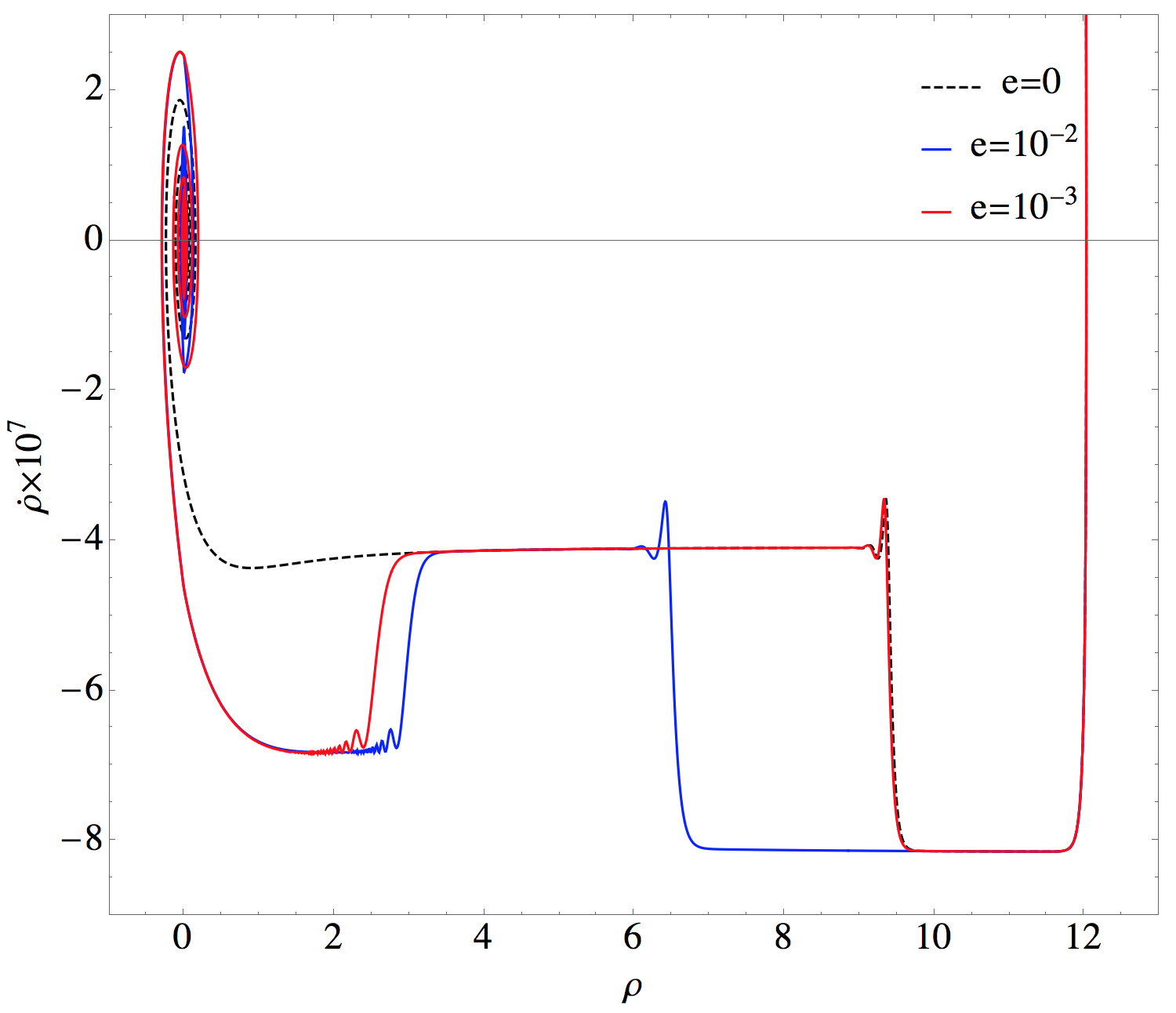}
		\caption{The phase space plot  with $c=2$  for different values of 
		 $\mathbf{e}=0, 10^{-2}$ and $10^{-3}$ from top to bottom, respectively. 
		As in left plot,  $m=10^{-6}M_{P}, \rho(0)=12M_{P}$ and $\dot \rho(0)=0$.}
		\label{cf2}
	\end{subfigure}
\end{figure}

To make the  qualitative behavior of the model more transparent, in Figure \ref{cf1} we have plotted the phase space diagram of $(\rho,\dot \rho)$ for the fixed value of $\mathbf{e}=0.01$ and for different values of $c=2, 2.2$ and $2.5$. As it is clear from this figure the starting time of the attractor phase depends on the value of $c$ parameter.  The larger values of $c$ correspond  to pumping more energy into the gauge field and, consequently, the attractor phase happens earlier. In Figure \ref{cf2}, the phase space diagram of $(\rho,\dot \rho)$ for the fixed value of $c=2$ but for different values of $\mathbf{e}=0,10^{-2}$ and $10^{-3}$ are plotted. As can be seen from this figure, the role of $\mathbf{e}$  becomes important only at the final stage of inflation. This is because  the gauge field scales like $A \propto 1/\tau^3 $ (see Eq. (\ref{A-c}) )
so the induced mass term $\mathbf{e}^2 A_\mu A^\mu \rho^2$ becomes important only towards the end of inflation when the inflaton field become massive, violating the slow-roll conditions.


\section{Perturbation Analysis}\label{pert}

In this section, we present the  cosmological perturbations of the model. For the purpose of Schwinger pair creation,  we are interested only in the scalar perturbations.

The scalar perturbations around the background geometry (\ref{FRW-t}) are given by
\begin{eqnarray}\label{metric-perturbation}
ds^{2} = - a^{2}(1+2N_1) d\tau^{2} + 2 a B_{,i} d\tau dx^i 
+ a^2 \big((1+2\psi)\delta_{ij}+E_{,ij}\big) dx^{i}dx^{j}\,,
\end{eqnarray}
where $N_1,B,\psi$, and $E$ are the scalar modes.

Scalar perturbations in the matter sector are given by
\begin{eqnarray}\label{matter-perturbations}
\varphi\rightarrow \rho+\delta \varphi\,, \hspace{1.5cm} 
A_{\mu} \rightarrow A_{\mu}+\delta A_{\mu}, \hspace{.3cm} \text{with} \hspace{.3cm}
\delta{A}_{\mu} = ( \delta A_{0}, \delta{A}_{x}, \partial_{y}M, 0 ) \,,
\end{eqnarray}
where $\delta \varphi, \delta{A}_0,\delta{A}_x$, and $M$ are the scalar modes in the matter sector.

In total, we have nine  scalar modes. However, not all of these modes are physical degrees of freedom. The spacetime diffeomorphism invariance fixes two scalar modes. We choose to work in the spatially flat gauge with 
\begin{equation}\label{SFG}
\psi = E = 0 \,.
\end{equation}
Moreover, the $U(1)$ invariance of the matter sector fixes one of the scalar modes. We work in Coulomb gauge  $\partial^{i} \delta A_{i}=0$, which implies
\begin{eqnarray}\label{Coulomb-gauge}
\partial^{x} \delta A_{x}=-\partial^{y}\partial_{y}M.
\end{eqnarray}

From the above relation, one of the scalar modes can be solved in terms of the other. Fixing the gauges, then we have  six scalar modes. The direct calculations show that the modes $N_1$, $B$, and $\delta{A}_0$ are non-dynamical which can be solved from a set of algebraic equations of motion and then substituted back into the quadratic action of the remaining perturbations. As shown in \cite{Emami:2013bk, 2014JCAP...08..027C} these non-dynamical modes are slow-roll suppressed \cite{2014JCAP...08..027C} and their contributions can be neglected to leading orders in slow-roll parameters. Note that the perturbation analysis for the gauge fields and the charged inflaton without neglecting the non-dynamical constraint equations has been considered in \cite{Lozanov:2016pac} .

Going to the Fourier space and using the two-dimensional rotational symmetry in $y-z$ plane in the small anisotropy limit and considering ${\bf k} = (k_x,k_y,0)=k(\cos{\theta},\sin{\theta},0)$, from the Coulomb gauge condition Eq. (\ref{Coulomb-gauge}) we find $M=-i\frac{k_x}{k_y^2} \delta{A}_x$. After substituting this in Eq. (\ref{matter-perturbations}), we obtain  
\begin{eqnarray}\label{A-perturbations}
\delta{A}_{\mu} =\delta{A}_{x} \Big(0, 1, -\frac{k_{x}}{k_{y}}, 0 \Big) \,.
\end{eqnarray}

The gauge field excitations $\delta A_{\mu}$ can also be decomposed into the transverse mode $D_{1}$ and the longitudinal mode $D_{2}$,  which are related to $\delta A_{x}$ and $M$ as 
\begin{eqnarray}\label{D1-D2-def}
D_{1} &\equiv& \delta A_{x} - ik\cos{\theta}M, \\
D_{2} &\equiv& \cos{\theta} \delta A_{x} + ik \sin^2{\theta}M.
\end{eqnarray}
Using the Coulomb gauge (\ref{Coulomb-gauge}), it is easy to show that 
\begin{eqnarray}\label{D1-D2}
D_{1} = \frac{\delta A_{x}}{\sin^2\theta}, \hspace{1cm} D_{2} = 0 \,,
\end{eqnarray}
which shows that the longitudinal mode does not propagate  in the Coulomb gauge (\ref{Coulomb-gauge}).

In summary, neglecting the contributions of the non-dynamical modes, we have only three dynamical scalar modes $(\delta\varphi,\overline{\delta\varphi},D_1)$ which propagate in a quasi dS background. 

Implementing the following field redefinitions for the canonically normalized fields $q$ and $D$, 
\begin{eqnarray}\label{canonical-fields}
q_k \equiv a\,\delta\varphi_k \,, \hspace{1.5cm} D_{k} \equiv  f\sin\theta\, D_{1k} \,,
\end{eqnarray}
and after performing some integration by parts, the quadratic action to leading orders in terms of the  small parameters  $I$ and $\epsilon$ is obtained to be 
\begin{eqnarray}\label{S20}
S_{2} &=& \frac{1}{2}\int d^{3}k d\tau \bigg[
|q'_k|^2 - \Big(k^2 - \frac{2+6I}{\tau^2} 
+ \frac{3\epsilon}{\tau^4H^2} 
+ \frac{{\bf e}^2 I\epsilon}{3\tau^2H^2} \big(\frac{\tau_e}{\tau}\big)^4  \Big)|q_k|^{2} 
\\ \nonumber 
&+& \frac{3I}{\tau^2}(q_k^2+\bar{q}_k^2)
+ {D'}_k^2 - \Big( k^2 - \frac{2}{\tau^2}
+ \frac{2{\bf e}^2}{\tau^2H^2\epsilon} \big(\frac{\tau_e}{\tau}\big)^4 \Big) D_k^2 
\\ \nonumber
&-& 2 \sqrt{\frac{2I}{3}} \frac{\sin\theta}{\tau^2} 
(6-\frac{{\bf e}^2}{H^2}\big(\frac{\tau_e}{\tau}\big)^4)
(q_k + \bar{q}_k) D_k
+ 2 \sqrt{6I} \frac{\sin\theta}{\tau} (q_k + \bar{q}_k) D'_k
\bigg] \,,
\end{eqnarray}
where we have set $M_{P}=1$ and a prime denotes derivative with respect to the conformal time. To obtain the above  action, we have used $\rho = \sqrt{ \frac{2}{\epsilon}}$ and $A = \sqrt{\frac{I\epsilon}{3}} a  \frac{\tau_e^2}{\tau^2}$ which are obtained  from the background equations.

The complex scalar modes $\bar{q}_k$ and  $q_k$  are  responsible for the Schwinger pair production process in our formalism. The corresponding equation of motion for $q$ can be obtained from the action (\ref{S20}) as follows 
\begin{equation}\label{Sasaki-Mukhanov}
q''_k + \Big( k^2 -\frac{2}{\tau^2} + \frac{m^2}{H^2\tau^2} 
+ \frac{{\bf e}^2 E_0^2}{9H^4\tau^2} \Big(\frac{\tau_e}{\tau}\Big)^4 \Big) q_k
= - 4\sqrt{2} \frac{E_0}{\sqrt{\epsilon}H} \sin\theta 
\Big(\big( 1 - \frac{{\bf e}^2\tau_e^4}{6H^2\tau^4} \big) \frac{D_k}{\tau^2} 
- \frac{D'_k}{2\tau} \Big) \,.
\end{equation}

It is instructive to compare Eq. (\ref{Sasaki-Mukhanov}) with Eq. (2.13) of Ref. \cite{2014JHEP...10..166K}. The authors in Ref. \cite{2014JHEP...10..166K} looked at a test complex scalar field in a dS background in the presence of a constant background electric field. Here, however, the quantum fluctuations of the inflaton field are responsible for the Schwinger pair production process. Moreover, in Ref. \cite{2014JHEP...10..166K}, a constant electric field in a dS spacetime is imposed by hand while in our model the electric field is driven by the dynamics of the model.  Besides, in our inflationary setup,  the slow-roll conditions provide a quasi dS setup and the gauge kinetic coupling  $f(\rho)$ prevents the electric field  from being  diluted  during inflation.  Here,  a constant background electric field can be obtained via $f\propto a^{-2}$ (i.e. with $c\simeq 1$).  In this regard, our inflationary scenario is more natural with minimum  number of parameters  to study the Schwinger  process during an inflationary era.

Now let us elaborate more on the Mukhanov-Sasaki equation (\ref{Sasaki-Mukhanov}) to study the Schwinger process and  charged particle production  in our model.  Our Eq. (\ref{Sasaki-Mukhanov}) is different than the corresponding equation used in previous works ,  e.g.  Eq. (2.13)  of  \cite{2014JHEP...10..166K} and those in \cite{2018JCAP...10..023B,2018PhRvD..98f3534S,2018JCAP...02..018G,2018PhRvD..98j3512K,2018arXiv181009815Z}  studying the Schwinger effect in cosmological scenarios,   in two aspects: i) In our case the quantum fluctuations of the electric field, encoded  in the scalar mode $D_k$, source the quantum fluctuation of the complex field and therefore they indirectly contribute to the pair production process. ii) The last term in the left hand side of  Eq. (\ref{Sasaki-Mukhanov}) is the effects of the background field in pair production process which is proportional to $a^{6}$ while it was proportional to $a^{2}$ in Ref. \cite{2014JHEP...10..166K}. The later result is also included in Ref. \cite{2018JCAP...02..018G,2018PhRvD..98j3512K} where a complex test field is considered in the context  of anisotropic inflation \cite{2009PhRvL.102s1302W}. Note that in our model, the quantum fluctuations of the inflaton field  are responsible for the Schwinger process of generating  charged pair particles. 

In this paper, we have two types of charged pair production: those coming from the background electric field which are encoded in the last term in the left hand side of Eq. (\ref{Sasaki-Mukhanov}) and those  coming from the quantum fluctuations of the gauge field in the right hand side of Eq. (\ref{Sasaki-Mukhanov}). The first is similar to what usually arise in non-perturbative analysis of Schwinger effect (see for instance Eq. (2.13) of Ref. \cite{2014JHEP...10..166K}) while
the latter is a new type of  perturbative source for the 
 charged pair production. In previous studies, the last term in the left hand side of Eq. (\ref{Sasaki-Mukhanov}) which was imposed by hand in the strong electric field regime,   was responsible for the non-perturbative Schwinger mechanism  in a dS background. However, in our  analysis this term is small during most of the period of inflation. In other words we are working in the weak electric field regime.  
 
 Note that the last term in the left hand side of Eq. (\ref{Sasaki-Mukhanov}) is proportional to $\tau^{-6}$ and it dominates towards the end of inflation  $\tau\to0$.  So, there is a critical time $\tau_c$ beyond which  the perturbative approximation  breaks down and Eq. (\ref{Sasaki-Mukhanov})  is no longer applicable. Though we should consider the non-perturbative Schwinger effect after $\tau_c$  but the time range is very short and inflation ends quickly once we cross $\tau_c$. In order to find the critical time $\tau_c$, let us work with the real and imaginary parts of the complex field $q$:
\begin{eqnarray}\label{v-u-def}
q\equiv v+iu \,, \hspace{1cm} \bar{q}\equiv v-iu \,.
\end{eqnarray}
In terms of the new fields defined in  Eq. (\ref{v-u-def}), the quadratic action (\ref{S20}) takes the following form
\begin{eqnarray}\label{S2}
S_{2} &=& \frac{1}{2}\int d^{3}k d\tau \bigg[
{u'}_k^2 - \Big( k^2 - \frac{2+3\epsilon}{\tau^2} + \frac{3\epsilon}{\tau^4H^2} 
+ \frac{\e^2 I\epsilon}{3\tau^2H^2} \big(\frac{\tau_e}{\tau}\big)^4 \Big) u_k^{2} 
 \nonumber \\
&+& {v'}_k^2 - \Big( k^2 - \frac{2+3\epsilon+12I}{\tau^2} + \frac{3\epsilon}{\tau^4H^2} 
+ \frac{\e^2 I\epsilon}{3\tau^2H^2} \big(\frac{\tau_e}{\tau}\big)^4 \Big) v_k^{2}
 \nonumber \\ 
&+& {D'}_k^2 - \Big( k^2 - \frac{2}{\tau^2}
+ \frac{2\e^2}{\tau^2H^2\epsilon} \big(\frac{\tau_e}{\tau}\big)^4 \Big) D_k^2 
 \nonumber  \\ 
&-& 4 \sqrt{6I} \sin\theta \Big( \frac{2D_k}{\tau^2} - \frac{D'_k}{\tau}
+ \frac{\e^2}{3H^2} \big(\frac{\tau_e}{\tau}\big)^4 \frac{D_k}{\tau^2} \Big) v_k
\bigg] \,.
\end{eqnarray}
One advantage of working with these new variables is that the quantum fluctuations of the electric field only couples with $v$ since $q+\bar{q}=2v$ and the imaginary component $u$ decouples completely.

The equations of motion for the modes $v$ and $D$ can be obtained from the action (\ref{S2}) as follows 
\begin{equation}\label{npv}
v''_k + \Big(k^2 -\frac{2+12I}{\tau^{2}} + \frac{3\epsilon}{H^{2}\tau^{4}}
+ \frac{I\epsilon {\e^2} M^2_P}{3 H^{2}\tau^2} \big(\frac{\tau_e}{\tau}\big)^4 \Big) v_k 
= 2\sqrt{6I}\sin{\theta}\Big( \frac{2D_k}{\tau^{2}} - \frac{D'_{k}}{\tau}
-\frac{\e^2 M^2_P}{3H^2} \big(\frac{\tau_e}{\tau}\big)^4 \frac{D_k}{\tau^{2}} \Big) \,,
\end{equation}
and 
\begin{equation}\label{npd}
D''_k + \Big(k^2-\frac{2}{\tau^{2}} +\frac{2 \e^2 M^2_P}{\epsilon H^2\tau^2} 
\big(\frac{\tau_e}{\tau}\big)^4 \Big)D_{k}=
2\sqrt{6I}\sin{\theta}\Big( \frac{v_k}{\tau^{2}} + \frac{v'_{k}}{\tau}
-\frac{\e^2 M^2_P}{3H^2} \big(\frac{\tau_e}{\tau}\big)^4
\frac{v_k}{\tau^{2}} \Big) \,.
\end{equation}

We will estimate the critical time $\tau_c$ as the time in which the system of coupled equations (\ref{npv}) and (\ref{npd}) can not be treated perturbatively. In order to do this, we should compare the interaction terms with each other in both equations. Let us first look at the mode $D$ in Eq. (\ref{npd}). There are two types of interaction terms. The self-interaction term containing $\frac{2 \e^2 M^2_P}{\epsilon H^2\tau^2} \big(\frac{\tau_e}{\tau}\big)^4$ in the left hand side of  Eq. (\ref{npd}) and the interaction with the mode $v$ in the right hand side of Eq. (\ref{npd}) giving rise to  $2\sqrt{6I}\sin{\theta}  \frac{\e^2 M^2_P}{3H^2\tau^2} \big(\frac{\tau_e}{\tau}\big)^4$. Comparing these terms with each other, and noting that  $\sqrt{\frac{2I}{3}} \epsilon  \ll 1$,  we find  that the self-interaction  term dominants earlier. Therefore, the critical time at which the mode $D$ starts to show non-perturbative behaviour is determined by the self-interaction term, given by
\begin{equation}\label{tauD}
\tau_{D}=\tau_{e} \Big( \frac{\e^2 M_P^2}{\epsilon H^2} \Big)^\frac{1}{4}= \tau_{e}
\Big( \frac{\e^{2}}{ 8\pi^2\epsilon^2{\cal P}_R} \Big)^\frac{1}{4} \,,
\end{equation}
where ${\cal P}_R = \frac{H^2}{8\pi^2M^{2}_{P}\epsilon} \sim 2.1 \times 10^{-9}$ is the curvature perturbation power spectrum.

In the same manner, for the mode $v$ we should  compare the two types of interactions  in Eq. (\ref{npv}) such as $\frac{I\epsilon {\e^2} M^2_P}{3 H^{2}\tau^2} \big(\frac{\tau_e}{\tau}\big)^4$ and
$2\sqrt{6I}\sin{\theta}\frac{\e^2 M^2_P}{3H^2} \big(\frac{\tau_e}{\tau}\big)^4$. The ratio of these interaction terms is $\sqrt{\frac{8}{3}}\frac{\sin\theta}{\sqrt{I}\epsilon}$. Therefore we  define 
the critical time associated to these interactions as
\begin{equation}\label{tauv}
\tau_{v} = \tau_{e} \Big( \frac{I \e^{2}}{48\pi^2{\cal P}_R} \Big)^\frac{1}{4} \,. 
\end{equation}
From Eqs. (\ref{tauv}) and (\ref{tauD}), we find
\begin{eqnarray}
\tau_{v}=\tau_{D} \Big(\frac{I \epsilon^2}{6}\Big)^\frac{1}{4} \,. 
\end{eqnarray}

As discussed before, the anisotropic inflationary models generate statistical anisotropies which are tightly constrained by the CMB data \cite{Ade:2015lrj}.  In order not to generate large statistical anisotropy we require $I \lesssim 10^{-7}$ \cite{Emami:2010hy,Emami:2013bk,Kanno:2010ab, Abolhasani:2013zya}. In addition, in order for the tensor perturbations to be perturbatively under control, we require ${\bf e} \lesssim 10^{-3}$ \cite{ 2014JCAP...08..027C}.  Taking the slow-roll parameter to be  $\epsilon\sim 10^{-2}$, we find (note that during inflation $\tau<0$)
\begin{eqnarray}\label{np-time}
\tau_D < \tau_e \lesssim  \tau_v \, .
\end{eqnarray}

This is an interesting result indicating  that we can study the mode $v$ perturbatively from the past infinity to the end of inflation $(-\infty,\tau_e]$,  while the mode $D$ becomes non-perturbative near the end of inflation. On the other hand, the mode $v$ is responsible for the charged pair production. This analysis shows that the non-perturbative Schwinger pair production does not take place during most of period of
inflation and may be relevant only towards the end of inflation. This conclusion is the key difference of our model compared to other less realistic scenarios  studied in previous works on Schwinger mechanism during inflation. This is because we took the complex scalar field to be the inflaton field itself which is responsible for curvature perturbations, and not a  hypothetical test field decoupled from inflation.


\section{ Pair Production}\label{pair}

In this section, we quantize the fields and find the number of charged pairs which are produced perturbatively during inflation. 

From the action (\ref{S20}) we obtain the conjugate momenta for the charged quantum fluctuations as $\Pi_q = \bar{q}'$ and $\bar{\Pi}_q = q'$. Then promoting $q$ to quantum operator and expanding them in terms of the mode functions, we have 
\begin{equation}\label{qdf}
\hat{q}(\tau,{\bf x}) = \int \frac{d^3k}{(2\pi)^3} \big(
a_{-{\bf k}} q_k(\tau) + {b}^{\dagger}_{\bf k} {\bar q}_{-k}(\tau) \big) e^{-i{\bf k.x}} \,,
\end{equation}
in which the  mode function $q_k$ satisfies Eq.  (\ref{Sasaki-Mukhanov}) accordingly.

Demanding the commutation relations $[\hat{q}(\tau,{\bf x}),\hat{\Pi}(\tau,{\bf x}')]= i \delta^{(3)}({\bf x}-{\bf x}')$ while  all other commutation relations being zero, we find the following well-known commutation relations between the annihilation and creations operators
\begin{equation}
[a_{\bf k},a^\dagger_{\bf k'}] = [b_{\bf k},
b^\dagger_{\bf k'}] 
= (2\pi)^{3} \delta^{(3)}({\bf k}-{\bf k}') \,.
\end{equation}

In the same manner, we quantize the quantum fluctuation of gauge field as 
\begin{equation}\label{D}
\hat{D}(\tau,{\bf x}) = \int \frac{d^3k}{(2\pi)^3} \big(c_{-{\bf k}} D_{k}(\tau)  
+ c^{\dagger}_{\bf k} {\bar D}_{k}(\tau)\big) e^{-i{\bf k.x}} \,,
\end{equation}
with 
\begin{equation}
[c_{\bf k},c^\dagger_{\bf k'}] = (2\pi)^{3} \delta^{(3)}({\bf k}-{\bf k}') \,,
\end{equation}
in which  the mode function $D_{k}$ satisfies Eq. (\ref{npd}). 

The correlation functions (power spectra) for the free parts of the modes $v,u$, and $D$ can be simply obtained if we neglect the effects of gauge field in background dynamics 
by setting $I=0$. More precisely, we can solve the  mode functions $v(\tau)$, $u(\tau)$ and $D(\tau)$ for the limit $I=0$. Then we take into account the effects of the gauge field on the inflaton field and the pair production  perturbatively  in terms of small parameters $I$ and $\e$. 

Before going further it is useful to express $u$ and $v$ in terms of the annihilation and creation operators. Substituting Eq. (\ref{qdf}) in definition (\ref{v-u-def}), we find
\begin{eqnarray}
\hat{v}(\tau,{\bf x}) = \int \frac{d^3k}{(2\pi)^3} \hat{v}_{k}(\tau) e^{-i{\bf k.x}}
= \frac{1}{2} \int \frac{d^3k}{(2\pi)^3} \big[(a_{-{\bf k}}+b_{-{\bf k}}) q_k(\tau) 
+ ({b}^{\dagger}_{\bf k}+{a}^{\dagger}_{\bf k}) {\bar q}_{k}(\tau) \big] e^{-i{\bf k.x}} \,,
\end{eqnarray}
and
\begin{eqnarray}
\hat{u}(\tau,{\bf x})
= \int \frac{d^3k}{(2\pi)^3} \hat{u}_{k}(\tau)e^{-i{\bf k.x}} = 
\frac{1}{2i} \int \frac{d^3k}{(2\pi)^3} \big[(a_{-{\bf k}}-b_{-{\bf k}}) q_k(\tau) + 
({b}^{\dagger}_{\bf k}-{a}^{\dagger}_{\bf k}) {\bar q}_{k}(\tau) \big] e^{-i{\bf k.x}} \, .
\end{eqnarray}

In order to implement the perturbative analysis, we decompose the quadratic action (\ref{S2}) to the free and interaction parts. Correspondingly,  the  interaction Hamiltonians are obtained to be 
\begin{eqnarray}
\label{Hint-uu}
H^{\rm int}_{uu} &=& -\frac{3}{2} I\epsilon \int \frac{d^3k}{(2\pi)^3}
\Big( 1 -\frac{\e^2}{9H^2} \big(\frac{\tau_e}{\tau}\big)^4 \Big) \frac{\hat{u}_{k}^2}{\tau^2} \,,
\\ 
\label{Hint-vv}
H^{\rm int}_{vv} &=& -\frac{3}{2} I \int \frac{d^3k}{(2\pi)^3} \Big( \epsilon + 4\cos{2\theta}
- \frac{\epsilon \e^2}{9H^2} \big(\frac{\tau_e}{\tau}\big)^4 \Big) \frac{\hat{v}_k^2}{\tau^2} \,,
\\
\label{Hint-vD}
H^{\rm int}_{vD} &=& 2\sqrt{6I} \sin\theta \int \frac{d^3k}{(2\pi)^3}
\Big( -\frac{2\hat{D}_k}{\tau}+\hat{D}'_k +
\frac{\e^2}{3H^2} \big(\frac{\tau_e}{\tau}\big)^4 \frac{\hat{D}_k}{\tau}\Big) \frac{\hat{v}_k}{\tau} \,,
\\
\label{Hint-DD}
H^{\rm int}_{DD} &=& \frac{1}{\epsilon}\int \frac{d^3k}{(2\pi)^3} \frac{\e^2}{H^2}
\big(\frac{\tau_e}{\tau}\big)^4 \frac{\hat{D}_k^2}{\tau^2} \,.
\end{eqnarray}
It is important to note that the interactions (\ref{Hint-uu}), (\ref{Hint-vv}), and (\ref{Hint-vD}) become large after the time $\tau_v$ defined in (\ref{tauv}) while from Eq. (\ref{np-time}) we see that they remain small till the end of inflation. The interaction (\ref{Hint-DD}), however, becomes large after the time $\tau_D$ given in Eq. (\ref{tauD}) and from Eq. (\ref{np-time}) it is clear that it can not be  treated perturbatively for $\tau\in(\tau_D,\tau_e]$. Therefore, we should be careful about the  time interval in which the interaction term (\ref{Hint-DD}) plays  role in our subsequent analysis. It is worth mentioning that this interaction will not contribute to the pair production process and curvature perturbation power spectrum. It only contributes to the power spectrum of the isocurvature mode $D$.

Starting with the Bunch-Davies initial condition,  the free mode functions are given by
\begin{equation}\label{BDS}
v_{k}=u_k=D_k
=\frac{i e^{-ik\tau}}{\sqrt{2k^{3}}\tau}\big( 1+ik\tau\big) \,.
\end{equation}

Now, we have all we need in hand to calculate the number of produced pairs. Before doing this,  it is useful to clarify the relation between the standard procedure of pair production by Schwinger mechanism through computing the Bogoliubov coefficients and the perturbative analysis that we apply here. It is well-known that a non-trivial background geometry, such as a time-dependent cosmological background, leads to a mixing between positive and negative frequency modes. In other words, the positive frequency mode function $q_{k}$ at asymptotic past $\tau\rightarrow-\infty$ is given by a linear combination of the positive and negative frequency modes at asymptotic future $\tau\rightarrow 0$. Choosing the vacuum $\ket{\Omega}$ in the asymptotic future by $\tilde{a}_{\bf k}\ket{\Omega}=\tilde{b}_{\bf k}\ket{\Omega}=0$, where
\begin{eqnarray}
\tilde{a}_{\bf k} = \alpha_{k}a_{\bf k}+{\bar \beta}_{k} b^{\dagger}_{-{\bf k}} \,, 
\hspace{1cm} 
\tilde{b}_{\bf k} = {\bar \beta}_{-{k}} a^{\dagger}_{-{\bf k}}+\alpha_{-{k}} b_{\bf k} \,,
\end{eqnarray}
with $\alpha_k$ and $\beta_k$ being the Bogoliubov coefficients, and assuming the usual commutation relations between $\tilde{a}_{\bf k}$ and $\tilde{a}_{\bf k}^{\dagger}$,  we have  $|\alpha_{k}|^{2}-|\beta_{k}|^{2}=1$. By simple algebra one can express $a_{\bf k}$ and $b_{\bf k}$ in terms of $\tilde{a}_{\bf k}$ and $\tilde{b}_{\bf k}$ as follows
\begin{eqnarray}
{a}_{\bf k}={\bar \alpha}_{k}\tilde{a}_{\bf k}-{\bar \beta}_{k}\tilde{b}^{\dagger}_{-{\bf k}}\,,
\hspace{1cm} 
b_{\bf k}={\bar \alpha}_{-{k}}\tilde{b}_{\bf k}-{\bar \beta}_{-{k}}\tilde{a}^{\dagger}_{-{\bf k}}.
\end{eqnarray}

\begin{figure}
\vspace{-10cm}
		\includegraphics[width=\textwidth]{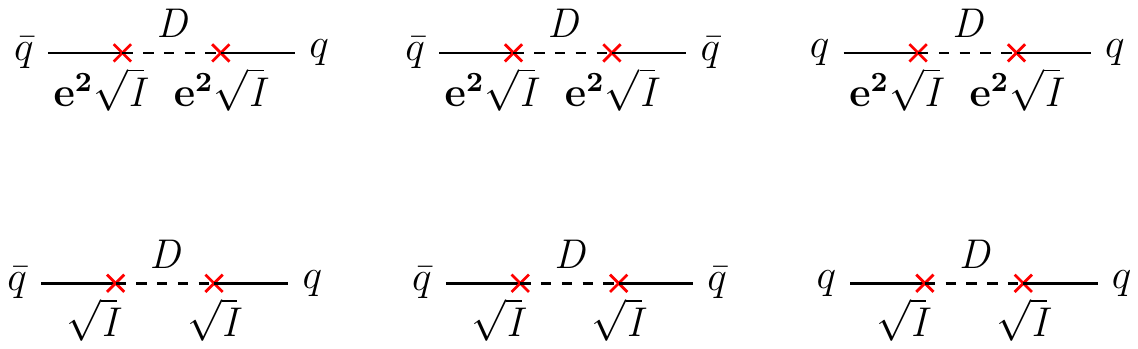}
\vspace{-10cm}
		\caption{Feynman diagrams corresponding to interaction Hamiltonian (\ref{Hint-vD}) giving rise perturbative pair production in (\ref{IN-IN-sub}).  These transfer vertices show the interactions between the charged scalar  field $q$ and its complex conjugate $\bar q$ with the gauge field excitation $D$.}\label{fig-feynmanvD}
\end{figure}

The number of the charged pairs produced from the infinite past to infinite future is 
\begin{eqnarray}\label{N}
{\cal N}_{k} = \big \langle \Omega | a^{\dagger}_{\bf k} a_{\bf k} |\Omega\big \rangle 
=\big \langle \Omega | b^{\dagger}_{-{\bf k}} b_{-{\bf k}} |\Omega\big\rangle = |\beta_{k}|^{2} \,.
\end{eqnarray}
The associated  Feynman diagrams giving non-zero contributions to Eq. (\ref{N})
are shown in Figure \ref{fig-feynmanvD}. In order to calculate ${\cal N}_{k}$ in Eq. (\ref{N}), we can implement the in-in formalism instead of computing the Bogoliubov coefficient through the  equation of motion (\ref{Sasaki-Mukhanov}). This is because  the interaction terms are small during  inflation, allowing for a perturbative in-in analysis.  

The in-in formula for the correlation functions of 
a typical scalar mode $\delta{X}$ is \cite{2005PhRvD..72d3514W,2010JCAP...04..027C}
\begin{eqnarray}\label{IN-IN}
\Delta \big \langle {\delta X^2 (\tau_c)} \big \rangle 
&=& \int_{\tau_{0}}^{\tau_{c}}d\tilde{\tau}_{1}\int_{\tau_{0}}^{\tau_{c}}
d\tau_{1} \langle0 | H_{I} (\tilde{\tau_{1}}) {\delta X^2 (\tau_c)} 
H_{I} (\tau_{1}) | 0 \rangle \nonumber \\
&-& 2 {\rm Re}\Big[ \int_{\tau_{0}}^{\tau_{c}}d\tau_{1} 
\int_{\tau_{0}}^{\tau_{1}}d\tau_{2} \langle0 |{\delta X^2 (\tau_c)} 
H_{I} (\tau_{1}) H_{I} (\tau_{2}) | 0 \rangle  \Big] + ... \,,
\end{eqnarray}
in which $H_{I}$ are the interaction Hamiltonians in the interaction picture and $| 0 \rangle $ is the free vacuum defined in the absence of the interactions. Looking at (\ref{N}),  in order to find the number of pairs, one can easily calculate  $\langle \Omega |a^{\dagger}_{\bf k}a_{\bf k'}| \Omega \rangle$ by means of 
Eq. (\ref{IN-IN}). The second term in the above formula  vanish since the effect of the creation operator on the free vacuum from the left vanishes and we simply have 
\begin{eqnarray}
\label{IN-IN-sub}
\bra{\Omega} { a^{\dagger}_{\bf k} a_{\bf k'}} \ket{\Omega} 
&=& \int_{-\infty}^{\tau_{e}}d\tilde{\tau}_{1} \int_{-\infty}^{\tau_e}d\tau_{1} \langle 0 | 
H^{\rm int}_{vD} (\tilde{\tau_{1}}) a^{\dagger}_{\bf k} a_{\bf k'} H^{\rm int}_{vD} (\tau_{1}) | 0 \rangle
\equiv  {\cal N}_k (2\pi)^3\delta^3({\bf k}-{\bf k'}) \,,
\end{eqnarray}
with 
\begin{eqnarray}\label{Nk}
{\cal N}_k = 
\bigg(\frac{3}{2} -\frac{4 \e^{2}M^{2}_{P}k^{4}\tau^{4}_{e}}{35H^{2}}
+\frac{8 \e^{4}M^{4}_{P}k^{8}\tau^{8}_{e}}{3675H^{4}}\bigg)
I\sin^2\theta N_k^2 \,,
\end{eqnarray}
where $N_k=-\ln(-k\tau_e)$ is the number of e-folds counted from the time when the mode of interest $k$ has left the horizon till the time of end of inflation. Note that the integrals are taken from the past infinity to the end of inflation.  This is because the  interaction Hamiltonian $H^{\rm int}_{vD}$  becomes non-perturbative only after $\tau_v \gtrsim \tau_e$ and therefore we are allowed to  perform the integrals from the past infinity to the end of inflation. 

The scale dependent of ${\cal N}_k$ is clear from  the last two terms  in  bracket in Eq. (\ref{Nk}).  To quantify this more appropriately, let us define  the scale $k_D$ for modes which leave the horizon at  $\tau= \tau_D$, corresponding to $k_D \tau_D  =-1$. Then using the expression for $\tau_D$ given in Eq. (\ref{tauD}), we obtain
\begin{eqnarray}
\label{Nk2}
{\cal N}_k = \bigg[ \frac{3}{2} -\frac{4 \epsilon}{35}  \left( \frac{k}{k_D} \right)^4
+\frac{8  \epsilon^2}{3675}  \left( \frac{k}{k_D} \right)^8  \bigg] I\sin^2\theta N_k^2 \, .
\end{eqnarray}
For observable CMB scales where $k \tau_e \sim k/k_D \rightarrow 0$, the dominant term in ${\cal N}_k $ is the first term  in  the bracket in Eq. (\ref{Nk}). In next Section, we relate this to the amplitude of quadrupolar statistical anisotropy. On the other hand, for modes which leave the horizon only towards the end of inflation corresponding to $k \gtrsim k_D$, ${\cal N}_k $ can start to grow. However, this period is short and inflation ends quickly afterwards. In addition, these scales are exponentially small compared to observable scales and can not have any interesting observables effects.  In a sense, the possible non-perturbative pair creations for scales 
smaller than $k_D^{-1}$ becomes entangled with preheating mechanism of particle creations on small scales at the end of inflation. 

In addition, we see that the number density of produced pairs is anisotropic, being proportional to $\sin^2\theta$ where $\theta$ is the angle between  mode number $\bfk$ and a preferred direction in the sky (the direction of anisotropy determined by the background electric field, which in our case is along the $x$-direction).

\section{Curvature and Isocurvature Power Spectra}\label{power}

In this section, we calculate the power spectra of the scalar modes in our model. We deal with a multiple field scenario with the scalar modes $u$ and $v$ coming from the quantum fluctuations of the complex inflaton field while the scalar mode $D$ coming from the quantum fluctuation of the electric field.

The comoving curvature perturbation is given by ${\cal R} =-\psi+ H\delta{u}$ where $\delta{u}$ is the velocity potential which is defined as $\delta T^t_i=\partial_i \delta{u}$. In the spatially flat gauge (\ref{SFG}), the curvature perturbation reduces to ${\cal R} = H\delta{u}$ with
\begin{equation}\label{velocity-potential}
\delta{u} = - \frac{q+\bar{q}}{2\phi'}= -\frac{v}{\phi'} \,.
\end{equation}
The above result shows that only the real part of the quantum fluctuations of the inflaton contribute to the curvature perturbations, yielding 
\begin{equation}\label{CP}
{\cal R} = H \delta{u} = -\Big(\frac{H}{\dot{\phi}}\Big)\frac{v}{a} \,.
\end{equation}
The associated two-point correlation function is given by
\begin{eqnarray}\label{CP-PS0}
\langle {{\cal R}_{\bf k}^\dagger {\cal R}_{\bf k'}} \rangle 
= \Big(\frac{H}{\dot{\phi}}\Big)^2 \frac{\langle{v^\dagger v}\rangle}{a^2}
= \frac{2\pi^2}{k^3} {\cal P}_{\cal R} (2\pi)^3 \delta({\bf k}-{\bf k}') \,,
\end{eqnarray}
where ${\cal P}_{\cal R}$ is the dimensionless power spectrum of the curvature perturbations.

To  find the power spectrum,  first we need to find the two-point correlation function for the real part of the scalar mode $v$ which is given by
\begin{equation}\label{CF-vv}
\langle{v_{\bf k}^\dagger v_{\bf k'}}\rangle = 
\frac{H^2}{2k^3} (2\pi)^3\delta({\bf k}-{\bf k}') + 
\Delta \langle{v_{\bf k}^\dagger v_{\bf k'}}\rangle \,,
\end{equation}
where the first term is obtained  using  Eq. (\ref{BDS}) coming from the free theory in the absence of interaction between the gauge field and the inflaton field. The second term in (\ref{CF-vv}), $\Delta \langle{v_{\bf k}^\dagger v_{\bf k'}}\rangle$,  represents contributions from the interactions listed in Eqs. (\ref{Hint-uu}), (\ref{Hint-vv}), (\ref{Hint-vD}) and (\ref{Hint-DD}).  The leading contributions in $\Delta \langle{v_{\bf k}^\dagger v_{\bf k'}}\rangle$   come  from the off-diagonal interaction Eq. (\ref{Hint-vD}) which are shown in the Feynman diagrams in Fig. \ref{fig-feynmanvD}, yielding
\begin{eqnarray}\label{correction-vv}
\Delta\langle{v_{\bf k}^\dagger v_{\bf k'}} \rangle 
=\frac{{H}^{2}}{k^{3}} \bigg(1 - \frac{\e^{2}}{42H^2N_k} 
+ \frac{\e^4}{4312H^4 N^{2}_k} \bigg) 
12I\sin^{2}{\theta}N^{2}_k (2\pi)^3 \delta^3({\bf k}-{\bf k'}) \,.
\end{eqnarray}

Substituting Eq. (\ref{correction-vv}) into Eqs. (\ref{CF-vv}) and  Eq. (\ref{CP-PS0}) we find the  curvature perturbations power spectrum  to be 
\begin{eqnarray}\label{CP-PS}
{\cal P}_{\cal R} = \frac{H^2}{8\pi^2\epsilon M_P^2} 
\Big(1 + 24  F(\beta) 
I \sin^{2}{\theta}N^{2}_{k}\Big) \,,
\end{eqnarray}
where we have defined the dimensionless parameter $\beta$  and the function $F(\beta)$ as follows
\begin{eqnarray}\label{beta}
\beta \equiv  \frac{\e^2 M_P^2}{42 H^2N_k} \, ,  \quad \quad 
F(\beta) \equiv 1 - \beta +\frac{9}{22} \beta^2 \, .
\end{eqnarray}
The  above result coincides with  the result obtained in \cite{2014JCAP...08..027C} using a different gauge.  Assuming $M_P/H \sim 10^{5}$ in chaotic inflation model, we would have $\beta \gtrsim1$ for $\e \gtrsim 10^{-4}$. However, as shown  in \cite{2014JCAP...08..027C}, in order to keep the anisotropies in tensor sector  under perturbative control  one actually requires $\e \lesssim 10^{-3}$, so in practice $\beta $ is not much bigger than unity. 

Conventionally, the statistical anisotropies in curvature perturbation power spectrum can be parameterized in terms of quadrupole amplitude $g_*$, defined via
\ba
{\cal P}_{\cal R} = {\cal P}_{\cal R}^{(0)} ( 1+ g_* \cos^2\theta) \, ,
\ea
where ${\cal P}_{\cal R}^{(0)}$ is the isotropic power spectrum in the absence of gauge field. Comparing the above definition  with our result obtained in Eq. (\ref{CP-PS}), we have
\ba
\label{g-star}
g_*  = -24 I N_k^2 F(\beta) \, .
\ea
There are tight observational constraints on the amplitude of $g_*$, requiring  $| g_*| < 10^{-2}$  \cite{Ade:2015lrj}. With $N_k\simeq 60$ for CMB scales, we conclude that $I \lesssim 10^{-7}$. 

We can now relate the number density of the created pairs to the amplitude of quadrupole anisotropy $g_*$. Using Eqs. (\ref{Nk2}) and (\ref{g-star}), and assuming  $\beta \sim {\cal O}(1)$, 
we find ${\cal N}_k \sim g_*$. This is an interesting result, providing a direct link between the amplitude of quadrupole anisotropy and the number density of the created pairs. As discussed before, the conventional non-perturbative Schwinger mechanism will not take place during inflation and it is only the perturbative pair creation which operates during most of the period of inflation. However, the key difference in our model compared to previous works is that  the complex scalar field is the inflaton field itself which at the same time is responsible for curvature perturbation. As a result, there is not much room for the efficiency of charged pair creation. This is  unlike other hypothetical setups where the complex scalar field assumed to be a test field decoupled from the inflationary sector and also where the electric field was given as a background field with no dynamical mechanism for its generation.

\begin{figure}
\vspace{-8cm}
		\includegraphics[width=\textwidth,scale=4]{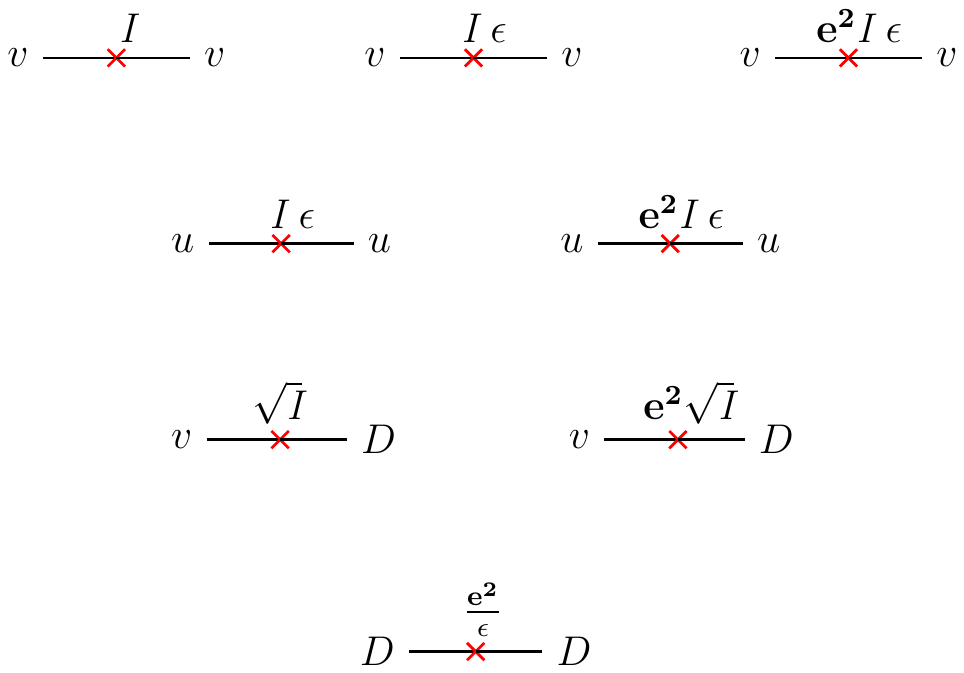}
\vspace{-8.5cm}
\caption{ The Feynman diagrams representing direct vertices originating from 
 the interaction Hamiltonians listed in Eqs. (\ref{Hint-uu}), (\ref{Hint-vv}), (\ref{Hint-vD}) and (\ref{Hint-DD}).}\label{ff3}
\end{figure}

To continue, we calculate the power spectra of the isocurvature modes which were not calculated in \cite{2014JCAP...08..027C}. With the quadratic action  (\ref{S2}) at hand we can easily calculate the power spectra of the other scalar modes as well. We note that the  scalar mode $u$ is completely decoupled from the other modes and therefore it is an isocurvature mode. One can directly solve the full equation of motion for the mode $u$ from Eq. (\ref{S2}). However, the  deviations from the standard free action are encoded in the interaction Hamiltonian (\ref{Hint-uu}) and it is easier to find the corrections through the in-in formula (\ref{IN-IN}). The corrections in the power spectrum of $u$ come only from the interaction Hamiltonian (\ref{Hint-uu}) with the Feynman diagrams shown in the second row of Figure \ref{ff3}. Performing the in-in integral, the correction in the power spectrum of $u$ is obtained to be 
\begin{eqnarray}
\label{correction-uu}
\Delta\langle{u_{\bf k}^\dagger u_{\bf k'}}\rangle 
= - \frac{{H}^{2}}{k^{3}} \bigg( 1 - 
\frac{\e^{2}}{84 H^2N_e}\bigg) I\epsilon N_k
(2\pi)^3 \delta^3({\bf k}-{\bf k'})\,.
\end{eqnarray}
Consequently, the corresponding dimensionless power spectrum  is
\begin{eqnarray}\label{CP-u}
{\cal P}_u = \Big(\frac{H}{2\pi}\Big)^2
\big(1 - (2-\beta) \big) I \epsilon N_k \,.
\end{eqnarray}

What remain is the scalar mode $D$, the fluctuation of the gauge field. This mode couples to the curvature perturbation through the off-diagonal interaction (\ref{Hint-vD}). Therefore, it is an entropy mode. Following Ref. \cite{Gordon:2000hv}, the entropy mode interacts with the curvature perturbations so we  should calculate their cross correlation. It is easy to see that the associated two-point function only receives contribution from the one-vertex Feynman diagrams between $v$ and $D$ which are shown in the third row of Figure \ref{ff3}. Correspondingly, the cross correlation between $v$ and $D$ modes is given by 
\begin{eqnarray}\label{correction-vD}
\Delta\langle{v_{\bf k}^{\dagger} D_{\bf k'}} \rangle 
= \frac{{H}^{2}}{k^{3}} \bigg(1-\frac{\e^{2}}{42{H}^{2}N_{k}} \bigg) 
\sqrt{6I}\sin{\theta} N_{k} (2\pi)^3 \delta^3({\bf k}-{\bf k'}) \, ,
\end{eqnarray}
yielding the dimensionless correlation function
\begin{eqnarray}\label{CP-vD}
{\cal P}_{vD} =  
2\big(1 - \beta \big) \sqrt{6I} \sin\theta N_k \,.
\end{eqnarray}

Finally,  we calculate the power spectrum of the mode $D$. For this purpose, we should be careful about the IR limit of the in-in integrals in  Eq. (\ref{IN-IN}) since the mode $D$ is perturbative only till  the time  $\tau= \tau_D$ given in Eq. (\ref{tauD}). The free wavefunction is given by the Bunch-Davies vacuum (\ref{BDS}) and is the same as the mode $v$ shown in Eq. (\ref{CF-vv}). The corrections in the power spectrum of $D$
come from the last Feynman diagrams in Fig. \ref{ff3}, yielding 
\begin{eqnarray}
\label{correction-DD}
\Delta\langle{D_{\bf k}^\dagger D_{\bf k'}}\rangle 
= - \frac{{H}^{2}}{k^{3}} \bigg( 1 +\frac{\e^{2}}{4H^2N_D}
- \frac{3 \e^4}{1232H^4N_D}
\bigg) \frac{8}{7} I\sin^{2}{\theta} N_D
(2\pi)^3 \delta^3({\bf k}-{\bf k'})\,,
\end{eqnarray}
where $N_D = - \ln(-k\tau_D)$.  The dimensionless power spectrum for the entropy mode $D$ then will be 
\begin{eqnarray}\label{CP-D}
{\cal P}_{D} = \Big(\frac{H}{2\pi}\Big)^2
\Big[1 - \frac{16}{7} \big(1+\frac{21}{2}\beta
-\frac{189}{44}\beta^2 N_D\big) I\sin^{2}{\theta}N_{D}\Big] \,.
\end{eqnarray}
We note that the contribution from the interaction Hamiltonian (\ref{Hint-DD}) becomes non-perturbative after the time $\tau_D$ while the mode $v$ is still  perturbative. The duration between $\tau_D$ and $\tau_e$, however, can not be too large since we need about $60$ e-folds of inflation to solve the flatness and the horizon problems. Note that this difference originates from the induced mass term $\e^2 A_\mu A^\mu \rho^2$. However, as argued before, this mass term should be negligible during most of the period of inflation to allow for slow-roll inflation. Consequently, the time difference between $\tau_e$ and $\tau_D$ is about  $1-2$ e-folds and we can practically set $N_D \simeq N_k$.


\section{Summary and Discussions}
\label{discussion}

In this work, we studied the efficiency of the Schwinger pair production in a minimal setup of inflation in which the inflaton field is a complex scalar field charged under the $U(1)$ gauge field. We have shown  that there are severe constraints on the efficiency of the Schwinger mechanism in this scenario.  In our setup the nearly constant electric field and a quasi-de Sitter background are natural attractor solutions of the field equations, in contrast to the previous considerations where a constant and uniform electric field in a fixed dS background geometry has been imposed by hand. Due to the smallness of the electric field energy density during much of the period of inflation, charged pair production could only occur perturbatively in this setup. The standard Schwinger pair production can become efficient only when the induced mass term $\e^2 A_\mu A^\mu \rho^2$ becomes significant. But in this limit a large effective mass is induced for the inflaton field which violates the slow-roll condition, ending inflation abruptly. Therefore, the non-perturbative Schwinger pair creation may take place only towards the final stages of inflation and on very small scales.

We have shown that the pair production in our inflationary model is negligible since the number of pairs in Eq. (\ref{Nk}) turned out to be proportional to the anisotropic parameter $I$ which is tightly constrained by the CMB observations, $I\lesssim 10^{-7}$. More specifically, we have shown that the number  of the created pairs is related to the amplitude of quadrupolar statistical anisotropy
$g_*$ which is tightly constrained by cosmological observations. One may wonder if the extension of  the setup to the case of isotropic model would yield significantly larger values of pair numbers. It is easy to see that this can not be the case. The isotropic extension of our model  was studied in \cite{2018arXiv181207464F} containing a triplet of $U(1)$ gauge field charged under 
complex scalar fields. There is no constraint from statistical anisotropy in the  isotropic extension of the current model. However, the condition that the curvature perturbation power spectrum to be nearly scale invariant requires that $I < 10^{-4}$. Although this is about three orders of magnitude larger than the bound on $I$ in our anisotropic model, but nonetheless  the number of created pairs are small. 

Recently it is shown that, in contrast to the Abelian U(1) case, when the Schwinger effect is driven by an SU(2) gauge field coupled to a charged scalar doublet,  both the Schwinger pair production and the induced current decrease as the interaction strength increases \cite{Lozanov:2018kpk,Maleknejad:2018nxz}. It is argued that  the isotropy of the SU(2) model plays a crucial  role in suppression of the particle production  rate and also the reduction of the induced current in the strong field limit.

It seems that if the inflaton field itself is a complex field (as in our model) and its quantum fluctuations to be responsible for the Schwinger pair production, then we can not achieve significant number of pairs.   One extension beyond our work is to apply the idea of quasi-single field inflation  \cite{2010JCAP...04..027C}. In this scenario the charged scalar field responsible for Schwinger pairs is a semi heavy charged scalar field while the inflaton field is a real scalar field coupled to the gauge field. This idea is a combination of \cite{2018JCAP...02..018G} and \cite{2018arXiv181009815Z}. Another option can be to look at Schwinger mechanism in the model of charge hybrid inflation \cite{Abolhasani:2013bpa}. In this model the inflaton field is a real scalar field while the complex scalar field is the waterfall field which  terminates inflation. The waterfall field is coupled to the $U(1)$ field which may lead to Schwinger pair production. In this setup, the sector responsible for generating  curvature perturbation (the inflaton field) is different than the sector responsible for the pair creation (the waterfall field), so the CMB constraints may be relaxed  and there may exist a corner of parameter space where the  Schwinger mechanism may be efficient. We plan to come back to this question in future. 

Since the  pair production in this minimal inflationary model is unmeasurably small on large (CMB) scales, therefore there is no need to consider the back-reaction effects of these particles. Consequently, the  Schwinger effect can not have any large scale effects such as constraining  inflationary magnetogenesis scenarios as envisaged in \cite{2014JHEP...10..166K}.

\vspace{1cm}


{\bf Acknowledgments:} We thank Jiro Soda for correspondences and comments on the draft and S. A. Hosseini Mansoori for assistance in xAct code.  M. A. Gorji thanks the Yukawa Institute for Theoretical Physics at Kyoto University for hospitality during the ``2019 YITP Asian-Pacific Winter School and Workshop on Gravitation and Cosmology" where this work was in its final stage.

{}

\end{document}